\documentclass[conference]{IEEEtranTCOM}
\normalsize
\newif\ifonecolumn
\onecolumnfalse
\usepackage{amsmath,amssymb}
\usepackage{float}
\usepackage{placeins}
\usepackage{graphicx}
\usepackage{amsmath}
\usepackage{amssymb}
\usepackage{color}
\usepackage{cite}
\usepackage{balance}
\usepackage{multirow}
\usepackage{array}
\usepackage{psfrag}
\usepackage{bm}

\newcommand*\diff{\mathop{}\!\mathrm{d}}

\newcommand*{\Scale}[2][4]{\scalebox{#1}{$#2$}}%

\renewcommand{\IEEEQED}{\IEEEQEDopen} %

\newcommand{\F}{\mathcal{F}}
\newcommand{\C}{\mathcal{C}}
\newcommand{\E}{\mathcal{E}}

\newcommand{\ftilde}{\tilde{f}}
\newcommand{\gtilde}{\tilde{g}}

\newcommand{\x}{\boldsymbol{x}}

\newtheorem{theorem}{Theorem}

\newtheorem{lemma}{Lemma}
\newtheorem{definition}{Definition}

\newtheorem{conjecture}{Conjecture}

\IEEEoverridecommandlockouts

\begin{document}
\title{Analysis of Spatially-Coupled Counter Braids}

\author{
\IEEEauthorblockN{Eirik Rosnes$^\dag$ and Alexandre Graell i Amat$^\ddag$}
\IEEEauthorblockA{$\dag$Department of Informatics, University of Bergen, N-5020 Bergen, Norway, and the Simula Research Lab.\\
  $\ddag$Department of Signals and Systems, Chalmers University of Technology, SE-412 96 Gothenburg, Sweden}\\
              \thanks{E.\ Rosnes was supported by the Norwegian-Estonian Research Cooperation Programme (grant EMP133) and by Simula@UiB. A.\ Graell i Amat was supported by the Swedish Research Council under grant \#2011-5961.}\vspace*{-1.0cm}
}

\maketitle

\begin{abstract}
A counter braid (CB) is a novel counter architecture introduced by Lu \emph{et al.} in 2007 for per-flow measurements on high-speed links. CBs achieve an asymptotic compression rate (under optimal decoding) that matches the entropy lower bound of the flow size distribution. Spatially-coupled CBs (SC-CBs) have recently been proposed. In this work, we further analyze single-layer CBs and SC-CBs using an equivalent bipartite graph representation of CBs. On this equivalent representation, we show that the potential and area thresholds are equal. We also show that the area under the extended belief propagation (BP) extrinsic information transfer curve (defined for the equivalent graph), computed for the expected residual CB graph  when a peeling decoder equivalent to the BP decoder stops, is equal to zero precisely at the area threshold. This, combined with simulations and an asymptotic analysis of the Maxwell decoder, leads to the  conjecture that the area threshold is in fact equal to the Maxwell decoding threshold and hence a lower bound on the maximum \emph{a posteriori} (MAP) decoding threshold. Finally, we present some numerical results and give some insight into the apparent gap of the BP decoding threshold of SC-CBs to the conjectured lower bound on the MAP decoding threshold.
\end{abstract}

\section{Introduction}

Recently, Lu \emph{et al.} proposed a novel counter architecture, inspired by sparse graph codes, for measuring network flow sizes, nicknamed Counter Braid (CB) \cite{lu08,lu08_1,lu07}. CBs use less memory space than other approximate counting techniques, since the flow sizes are compressed \textit{on-the-fly}. They are indeed asymptotically optimal (under some mild conditions) \cite{lu07}, i.e., the average number of bits needed to store the size of a flow tends to the information-theoretic limit (under maximum-likelihood decoding) when the number of flows goes to infinity. Furthermore, they are characterized by a layered structure which can be described by a graph. This allows for low-complexity decoding using a message passing (belief propagation (BP)) decoding algorithm. In general, good performance can be achieved with a small number of layers.

Spatially-coupled CBs (SC-CBs) were recently introduced by Rosnes in \cite{ros14}, with improved BP decoding thresholds as compared to uncoupled
CBs. It was numerically shown in \cite{ros14} that the BP decoding threshold of single-layer SC-CBs converges to a fixed value. 

In this paper, we further analyze single-layer CBs and SC-CBs. We consider an equivalent bipartite  graph representation of CBs, with identical finite-length performance and asymptotic behavior to that of CBs decoded on the original bipartite graph. On this equivalent representation, we show that the potential threshold, introduced by Yedla \emph{et al.} in \cite{yed13}, and the area threshold are equal. We also show that the area under the extended BP (EBP) extrinsic information transfer (EXIT) curve (defined for the equivalent graph), computed for the expected residual CB graph  when a peeling decoder equivalent to the BP decoder stops, is equal to zero precisely at the area threshold. This, combined with simulations and an asymptotic analysis of the Maxwell decoder, leads to the  conjecture that the area threshold is equal to the Maxwell decoding threshold and hence a lower bound on the maximum \emph{a posteriori} (MAP) decoding threshold.

Interestingly, when coupling the original (or the equivalent) graph, there is a  remaining gap between the potential threshold and the BP decoding threshold of SC-CBs even in the limit of large coupling chain length and smoothing parameter. This gap seems to be fundamental and due to the fact that the flow node update rule for even and odd iterations is different. 

In this text, all proofs are omitted due to lack of space. An extended version containing proofs will be published on arXiv.

\begin{figure}[!t]
\psfrag{f11}{$\Scale[0.5]{f^{(1)}_{1}}$}
\psfrag{f12}{$\Scale[0.5]{f^{(1)}_{2}}$}
\psfrag{f13}{$\Scale[0.5]{f^{(1)}_{3}}$}
\psfrag{f14}{$\Scale[0.5]{f^{(1)}_{m_0}}$}
\psfrag{f21}{$\Scale[0.5]{f^{(2)}_1}$}
\psfrag{f22}{$\Scale[0.5]{f^{(2)}_2}$}
\psfrag{f23}{$\Scale[0.5]{f^{(2)}_3}$}
\psfrag{f24}{$\Scale[0.5]{f^{(2)}_{m_1}}$}
\psfrag{c11}{$\Scale[0.5]{c^{(1)}_{1} = \Xi_{2}(f^{(2)}_1)}$}
\psfrag{c12}{$\Scale[0.5]{c^{(1)}_{2} = \Xi_{2}(f^{(2)}_2)}$}
\psfrag{c13}{$\Scale[0.5]{c^{(1)}_{3} = \Xi_{2}(f^{(2)}_3)}$}
\psfrag{c14}{$\Scale[0.5]{c^{(1)}_{m_1} = \Xi_{2}(f^{(2)}_{m_1})}$}
\psfrag{c21}{$\Scale[0.5]{c^{(2)}_{1}}$}
\psfrag{c22}{$\Scale[0.5]{c^{(2)}_{2}}$}
\psfrag{c23}{$\Scale[0.5]{c^{(2)}_{m_2}}$}
\psfrag{G1}{$\Scale[0.5]{\mathcal{G}_{1}}$}
\psfrag{G2}{$\Scale[0.5]{\mathcal{G}_{2}}$}
\centering
\includegraphics[width=\columnwidth,height=4.35cm]{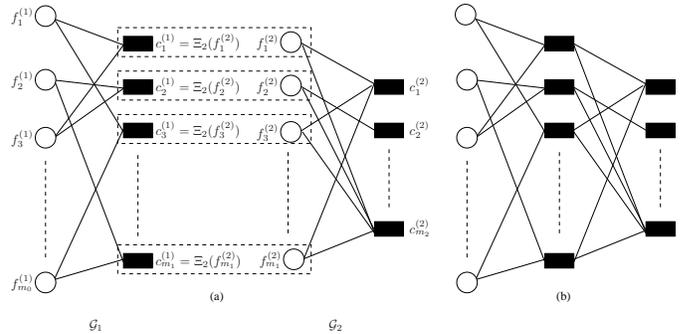}
\vspace{-4ex}
\caption{The graph of a two-layer CB. $(a)$ the bipartite graphs $\mathcal{G}_1$ and $\mathcal{G}_2$; $(b)$ the combined graph.}%
\label{fig:cb}
\vspace{-2ex}
\end{figure}

\section{Counter Braids}
\label{sec:CounterBraids}

A CB is a counter architecture consisting of $L \geq 1$ layers. At layer $l \in \{1,\ldots,L\}$, it has $m_l$ counters of depth $d_l$ bits, with $m_i<m_j$ for $i>j$. The number of distinct flows to be counted is denoted by $m_0$.

The $l$-th layer of a CB can be represented by a bipartite graph $\mathcal{G}_l = \mathcal{G}_l(\F_l \cup \C_l,\E_l)$, where $\F_l$ (of size $m_{l-1}$) denotes the set of flow nodes, and $\C_l$  (of size $m_{l}$) denotes the set of counter nodes. The set of edges of the graph is denoted by $\E_l$. The $l$-th layer is connected to the $(l-1)$-th layer by a bijective mapping $\Xi_l(f) = c$ on the set of flow nodes of the $l$-th layer, where $f \in \F_l$ and $c \in \C_{l-1}$, $l=2,\ldots,L$. For notational convenience, the neighborhood of a node $a$ (either a counter node $c$ or a flow node $f$) is denoted by $\delta(a)$.

When a flow is encountered (for instance, on a high-speed link), all connected counter nodes of the first layer are incremented modulo $2^{d_1}$. If a counter node $c$ of the first layer overflows, all connected counter nodes of the second layer (formally the counter nodes in the set $\delta(\Xi^{-1}_2(c))$), are incremented modulo $2^{d_2}$. Furthermore, if a counter node in the second layer overflows, all connected counter nodes of the third layer are incremented modulo $2^{d_3}$. This process is repeated for each level until we reach the final layer $L$. We denote by $\phi(c)$ the final value of a counter node $c$ prior to decoding, and by $\hat{\phi}(f)$ the estimated value (after decoding) of a flow node $f$.  %

An example of a two-layer CB is shown in Fig.~\ref{fig:cb}, where flow nodes are represented by empty circles and counter nodes by filled squares. Fig.~\ref{fig:cb}$(a)$ shows the bipartite graphs $\mathcal{G}_1$ and $\mathcal{G}_2$ of the two-layer CB, while Fig.~\ref{fig:cb}$(b)$ depicts an equivalent graph where a flow node of the second layer and its corresponding counter node of the first layer in Fig.~\ref{fig:cb}$(a)$ are represented using a single \textit{combined} counter node due to the bijective mapping $\Xi_2(\cdot)$.

In the following, we will assume that the bipartite graphs $\mathcal{G}_l$, $l=1,\ldots,L$, are left-regular, i.e., $|\delta(f)| = k_l$, for some integer $k_l \geq 2$, and for all $f \in \F_l$. The assignment of graph sockets of flow nodes to graph sockets of counter nodes, i.e., the connections in the graph, is done in a random fashion. This means that, asymptotically as the number of flow nodes tends to infinity (while $\frac{|\C_l|}{|\F_l|}$ is kept fixed), the distribution of the fraction of edges connected to a given %
counter node approaches a Poisson distribution \cite{lu08}. We can now formally define the flow and counter node degree distributions. For counter nodes, we assume the asymptotic Poisson distribution. Let 
$L_l(z)  = z^{k_l}$, $k_l \geq 2$, and %
\begin{align*}  
R_l(z) = \sum_{i=0}^{\infty} R^{(l)}_i z^i = \sum_{i=0}^{\infty} \frac{\mathrm{e}^{-\gamma_l} (\gamma_l z) ^{i}}{i!}
\end{align*}
denote the node-perspective flow node  and  counter node degree distributions of the $l$-th layer, respectively.  Here, $\gamma_l = \frac{m_{l-1}k_l}{m_l}$ is the average counter node degree. The corresponding edge-perspective  degree distributions are 
\begin{align}
\lambda_l(z) &= \frac{L_l'(z)}{L_l'(1)} = z^{k_l-1} \notag 
\end{align}
and
\begin{align}
\rho_l(z) &= \frac{R_l'(z)}{R_l'(1)} = \sum_{i=0}^{\infty} \frac{\mathrm{e}^{-\gamma_l} (\gamma_l z)^{i}}{i!} =  R_l(z), \notag 
\end{align}
respectively. Finally, we denote by $\beta_l = \frac{m_{l}}{m_{l-1}}$ the fraction of the number of counter and flow nodes of the $l$-th layer.

\subsection{Belief Propagation Decoding} \label{sec:message_passing}

CBs can be decoded using a BP decoding algorithm \cite{lu08} on the corresponding graph, starting with the right-most layer. After decoding the $l$-th layer, represented by the bipartite graph $\mathcal{G}_l$, the counter nodes of the $(l-1)$-th layer are updated based on the mapping $\Xi_{l}(\cdot)$. 

Consider the decoding of the $l$-th layer in more detail. We define the messages exchanged in the BP decoding algorithm as follows. Let $\mu^{(\ell)}_{f \rightarrow c} \in \mathbb{R}$, where $\mathbb{R}$ is the field of real numbers and $(f,c) \in \E_l$, denote the message sent from flow node $f$ to counter node $c$ for some $l$, during the $\ell$-th iteration of the BP decoding algorithm. Likewise, let $\psi^{(\ell)}_{c \rightarrow f} \in \mathbb{R}$, $(f,c) \in \E_l$, denote the message sent from counter node $c$ to flow node $f$ during the $\ell$-th iteration of the algorithm. The counter node and flow node update rules (for $\ell=1,\ldots,\ell_{\max}$) are given as follows\cite{lu08},
\begin{align}
\psi^{(\ell)}_{c \rightarrow f} &= \max \left\{ \phi(c) - \sum_{ f' \in \delta(c) \setminus f} \mu^{(\ell-1)}_{f' \rightarrow c},\;f_{\rm min} \right\} \label{eq:Bp1} \\
\mu^{(\ell)}_{f \rightarrow c} &= \begin{cases}
\min_{c' \in \delta(f) \setminus c} \psi^{(\ell)}_{c' \rightarrow f}, & \text{if $\ell$ is odd} \\
\max_{c' \in \delta(f) \setminus c} \psi^{(\ell)}_{c' \rightarrow f}, & \text{if $\ell$ is even} \end{cases}, \label{eq:Bp2}
\end{align}
where $f_{\rm min}$ is the minimum flow size (from the flow size distribution) of a flow node, $\ell_{\max}$ is the maximum number of iterations, and $\mu^{(0)}_{f \rightarrow c} = f_{\rm min}$ for all $(f,c) \in \E_l$. The final estimate of the flow sizes of the flow nodes is according to
\begin{displaymath}
\hat{\phi}(f) = \begin{cases}
\min_{c \in \delta(f)} \psi^{(\ell_{\max})}_{c \rightarrow f}, & \text{if $\ell_{\max}$ is odd} \\
\max_{c \in \delta(f)} \psi^{(\ell_{\max})}_{c \rightarrow f}, & \text{if $\ell_{\max}$ is even} \end{cases}. \notag
\end{displaymath}

After decoding the $l$-th layer (starting with the $L$-th layer), the counter nodes of the $(l-1)$-th layer are updated as follows (for all $f \in \F_{l}$),
\begin{displaymath}
\phi(\Xi_{l}(f)) \leftarrow \hat{\phi}(f) \cdot 2^{d_{l-1}} + \phi( \Xi_{l}(f)).
\end{displaymath}
Decoding proceeds layer-by-layer until the first layer is decoded. 

In the rest of the paper we assume a single-layer system and frequently omit, for notational convenience, the  subscript/superscript $l$. Also, we assume that the depth of the counters $d_1$ %
is large enough so that they do not overflow.

\section{Asymptotic Analysis of Single-Layer Counter Braids}
\label{sec:AsymptoticAnalysis}

Consider a single-layer CB with $m_0$ flow nodes and $m_1$ counter nodes of depth $d_1$. When $m_0\rightarrow \infty$, the asymptotic performance of CBs can be analyzed by means of density evolution (DE). Denote by $x^{(\ell)}$ and $y^{(\ell)}$ the error probability of an outgoing message from a flow node and a counter node, respectively, at the $\ell$-th iteration. The flow node and counter node DE updates at iteration $\ell$ are described by the equations
\begin{align}
\label{eq:DECounter1}
x^{(\ell)} = \ftilde\left(y^{(\ell)}; \epsilon\right),~~~~
y^{(\ell)} = \gtilde \left(x^{(\ell-1)}\right),
\end{align}
where
\begin{align}
\label{eq:DECounter1a}
 \ftilde(y;\epsilon) &= \begin{cases}
y^{k-1},  & \text{if $\ell$ is odd} \\
\epsilon \cdot y^{k-1}, & \text{if $\ell$ is even}\end{cases}\\
\label{eq:DECounter1b}
\gtilde(x) & = 1-\rho\left(1-x\right)
\end{align}
and $\epsilon$ is the probability of observing a flow of size strictly larger than $f_{\rm min}$. 

In the following, we denote by ${\cal{X}}=[0,1]$ and ${\cal{Y}}=[0,1]$ the set of possible values for $x$ and $y$, respectively, and by ${\cal{E}}$ the set of possible values for $\epsilon$.

For a given flow size distribution, or more precisely, for  a given $\epsilon$, a relevant parameter for the performance of single-layer CBs is the number of counters per flow, $\beta=\frac{m_1}{m_0}=\frac{k}{\gamma}$. The average number of bits needed to represent a flow is therefore $\beta d_1$, hence $\beta$ is directly related to the compression rate. In particular, we are interested in the minimum value of $\beta$ so that decoding is successful, 
\begin{align*}
\beta_{\rm BP} = \beta_{\rm BP}(\epsilon)\triangleq\inf\left\{\beta>0 \; |\; x^{(\infty)}(\beta,\epsilon)=0\right\}.
\end{align*}
Alternatively, we can analyze the asymptotic behavior of CBs by fixing $\beta$ and finding the maximum value of $\epsilon$ such that decoding is successful, 
\begin{align*}
\epsilon_{\rm BP} = \epsilon_{\rm BP}(\beta)\triangleq\sup\left\{\epsilon\in\mathcal{E} \; |\;  x^{(\infty)}(\beta,\epsilon)=0\right\}
\end{align*}
since for a fixed $k$ and any $\epsilon$ it follows that $\epsilon_{\rm BP}(\beta_{\rm BP}(\epsilon))=\beta_{\rm BP}^{-1}(\beta_{\rm BP}(\epsilon))=\epsilon$.
 In this case, $\epsilon_{\rm BP}$ has a similar meaning as the BP decoding threshold for low-density parity-check (LDPC) codes over the binary erasure channel, where $\epsilon$ can now be interpreted as the channel parameter.

Combining two successive iterations into a single one, 
we can write the DE in an equivalent form as
\begin{align}
\label{eq:DECounter2}
x^{(2\ell)} = f\left(y^{(2\ell)}; \epsilon\right),~~~~
y^{(2\ell)} = g \left(x^{(2\ell-2)}\right),
\end{align}
where
\begin{align}
f(y;\epsilon) &= \epsilon \cdot y^{k-1} \label{eq:f} \\
g(x) &= 1-\rho\left(1-\left(1-\rho\left(1-x\right)\right)^{k-1}\right). \label{eq:g}
\end{align}

Clearly, the DE recursions in (\ref{eq:DECounter1})-(\ref{eq:DECounter1b}) and (\ref{eq:DECounter2})-(\ref{eq:g}) give the same BP decoding threshold. %

There is a way to construct an equivalent graph representation of single-layer CBs (and a corresponding message passing decoding algorithm) which will give identical finite-length performance to that of the original graph, and asymptotically the DE recursion in (\ref{eq:DECounter2})-(\ref{eq:g}). The new graph has the same flow nodes as the original graph and the same number of counter nodes, but each counter node represents now a tree of depth three grown from the corresponding counter node in the original graph. The overall construction  is omitted due to lack of space. Thus, in the following, we can equivalently base our analysis on the recursion in  (\ref{eq:DECounter2})-(\ref{eq:g}).

\begin{lemma} \label{th:1}
The functions $f(y;\epsilon)$ (for a fixed value of $\epsilon > 0$) and $g(x)$, $k \geq 2$, are strictly increasing in $y$ and $x$, respectively.
\end{lemma}

As a result of Lemma~\ref{th:1}, it follows that the DE recursion in (\ref{eq:DECounter2})-(\ref{eq:g}) (and thus the DE recursion in (\ref{eq:DECounter1})-(\ref{eq:DECounter1b})) for uncoupled CBs converges to a fixed-point. The fixed-point DE equation for $x = x^{(\infty)}(\beta,\epsilon)$ is
\begin{align}
\label{eq:DECounter2FixedPoint}
x = f(g(x)).
\end{align}
Decoding is successful if the fixed-point is $x^{(\infty)}(\beta,\epsilon)=0$. 

We will need the definition of an \emph{admissible system} \cite{yed13}.

\begin{definition}[Admissible System] \label{lem:prop}
An admissible system is a system where the functions $f(y;\epsilon)$ and $g(x)$ satisfy the following properties.
\begin{enumerate}
\item The first derivative of $f(y;\epsilon)$ exists and is continuous on $\cal{Y}\times\E$, the first derivative of $g(x)$ exists and is continuous on $\cal{X}$,
\item $f(y;\epsilon)$ is nondecreasing in both $y$ and $\epsilon$, %
\item $g(x)$ is strictly increasing in $x$, and
\item the second derivative of $g(x)$ exists and is continuous on $\cal{X}$.
\end{enumerate}
\end{definition}

It is easy to show that the DE updates in (\ref{eq:f})-(\ref{eq:g}) satisfy these properties.%

\subsection{Extended Belief Propagation Extrinsic Information Transfer Curve}

The EBP EXIT curve of a single-layer CB is given in parametric form by
\begin{equation} \label{eq:EBP_EXIT}
(\epsilon,h^{\rm EBP}) = (\epsilon(x), (1-\rho(1-(1-\rho(1-x))^{k-1}))^{k}),
\end{equation}
where
\begin{equation} \notag %
\epsilon(x)= \frac{x}{(1-\rho(1-(1-\rho(1-x))^{k-1}))^{k-1}}
\end{equation}
is the solution of (\ref{eq:DECounter2FixedPoint}) for $\epsilon$,
and $x \in {\cal{X}}$. The curve is a trace of all fixed-points of the DE recursion in (\ref{eq:DECounter2FixedPoint}) (both stable and unstable).

\begin{definition}[Area Threshold] \label{def:MAP}
Let $(\epsilon(x^*), h^{\rm EBP}(x^*))$ be a point on the EBP EXIT curve $h^{\rm EBP}$ of a single-layer CB such that 
\begin{displaymath}
\int_{x^*}^1 h^{\rm EBP}(x)\, \diff \epsilon(x) = \int_{0}^1 h^{\rm EBP}(x)\, \diff \epsilon(x)
\end{displaymath}
and there exist no $x' \in (x^*,1]$ such that $\epsilon(x') = \epsilon(x^*)$. Then, the area threshold, denoted by $\bar{\epsilon}$, is defined as $\bar{\epsilon} = \epsilon(x^*)$. %
\end{definition}

\subsection{Potential Function, Potential Threshold, and Area Threshold}

Since the DE recursion in (\ref{eq:DECounter2})-(\ref{eq:g})   %
describes an admissible system (see Definition~\ref{lem:prop}), we can define a corresponding potential function as in \cite{yed13} for the uncoupled system.

\begin{definition}
The (uncoupled) potential function $U(x;\epsilon)$ of the system defined by the functions $f$ and $g$ from (\ref{eq:f}) and (\ref{eq:g}), respectively, is given by
\begin{displaymath}
\begin{split}
U(x;\epsilon) &\triangleq x g(x) - \int_{0}^x g(z)\,\diff z - \int_{0}^{g(x)} f(z;\epsilon)\,\diff z \\
&= x g(x) - \int_{0}^x g(z)\,\diff z - \frac{\epsilon}{k} g(x)^k.
\end{split}
\end{displaymath}
\end{definition}

Following \cite[Def.\ 32]{yed13}, we make the following definitions,
\begin{align}
\Psi(\epsilon) &\triangleq \min_{x \in \cal{X}} U(x;\epsilon), \notag \\
X^*(\epsilon) &\triangleq \{x \in {\cal{X}}\; |\; U(x;\epsilon) = \Psi(\epsilon) \}, \text{ and} \notag \\
\bar{x}^*(\epsilon) &\triangleq \max X^*(\epsilon). \notag
\end{align}
Now, we can define the \emph{potential threshold} $\epsilon^*_{\rm p}$ as \cite[Def.\ 35]{yed13}
\begin{equation} \label{eq:area_thres}
\epsilon^*_{\rm p} \triangleq \sup \left\{\epsilon \in {\cal{E}}\; |\;  \bar{x}^*(\epsilon) = 0\right\}.
\end{equation}

The following theorem shows that the area threshold and the potential threshold are equal, $\bar{\epsilon}=\epsilon^*_{\rm p}$.
\begin{theorem}
The area threshold from Definition~\ref{def:MAP} is equal to the potential threshold from (\ref{eq:area_thres}).
\end{theorem}

\section{Maxwell Decoder} \label{sec:Maxwell}

A Maxwell decoder \cite{mea08} can be constructed for CBs as for LDPC codes, and it can be analyzed asymptotically using DE on the equivalent graph representation mentioned in Section~\ref{sec:AsymptoticAnalysis}. An important point to consider is that the Maxwell decoder is in general \emph{not} a MAP decoder as for LDPC codes, since the flow size distribution is in general nonuniform. Thus, the \emph{Maxwell decoding threshold} (on $\epsilon$) is in general a lower bound on the \emph{MAP decoding threshold}. %

Now, note that the BP decoder with the update rules in (\ref{eq:Bp1})-(\ref{eq:Bp2}) will always \emph{stop}, i.e., at some point the flow size estimates of the flow nodes will be the same at iterations $\ell$ and $\ell-2$ for some $\ell$. When the decoder stops, either the flow size estimates for a flow node at iterations $\ell$ and $\ell-1$ are the same (i.e., we have convergence (upper and lower bounds are the same and thus we have the correct value for the flow size)), or we have oscillation (i.e., the estimates for iterations $\ell$ and $\ell-2$ are the same, but the estimates for iterations $\ell$  and $\ell-1$ are not the same).

The BP decoder can be turned into a peeling decoder in the following way. Run the BP decoder and in addition apply the following two peeling rules: 1) In case $|\delta(c)|=1$ for a counter node $c$, then remove $c$ and the connected flow node $f \in \delta(c)$ and all its edges from the graph. Decrease the values of the counter nodes of $\delta(f) \setminus c$ by the value of $c$. 2) For odd iterations, if a message from a counter node $c$ to a flow node $f$ is equal to $f_{\rm min}$, then remove the flow node $f$ and all its edges from the graph. Decrease the values of the counter nodes of $\delta(f)$ by $f_{\rm min}$. The BP and the peeling decoders are \emph{equivalent}, in the sense that the set of \emph{converged} flow nodes for the BP decoder is exactly equal to the set of \emph{peeled} flow nodes for the peeling decoder.

\begin{theorem} \label{th:maxwell_threshold}
The EBP EXIT curve (defined for the equivalent graph as in (\ref{eq:EBP_EXIT})) for the expected residual CB graph when the peeling decoder stops is given in parametric form by
\begin{displaymath}
(\tilde{\epsilon},\tilde{h}^{\rm EBP}) = (\tilde{\epsilon}(z;x), (1-\tilde{\rho}(1-z;x))^{k}),
\end{displaymath}
where
\begin{equation} \notag %
\tilde{\epsilon}(z;x)= \frac{z}{(1-\tilde{\rho}(1-z;x))^{k-1}} \text{ and } \tilde{\rho}(z;x) = 1 -\frac{g(x-zx)}{g(x)},
\end{equation}
and where $x=x(\epsilon)$ is the largest fixed-point of the DE recursion in (\ref{eq:DECounter2FixedPoint}), for a given $\epsilon$. Furthermore, for $\epsilon = \bar{\epsilon}$ (the area threshold from Definition~\ref{def:MAP}), the area $\int_{0}^1 \tilde{h}^{\rm EBP}(z;x) \, \diff \tilde{\epsilon}(z;x)$  is equal to zero.
\end{theorem}

Theorem~\ref{th:maxwell_threshold} combined with simulations (not included here) and the fact that an expected lower bound (analogous to Lemma~11 in \cite{mea08}) on the number of independent guesses (in the asymptotic limit) for a Maxwell decoder computed from its DE recursion on the equivalent graph is equal to zero exactly at the area threshold (details omitted for brevity), leads to the following conjecture.

\begin{conjecture} \label{th:MAP}
The area threshold from Definition~\ref{def:MAP}, $\bar{\epsilon}$, is equal to the Maxwell decoding threshold and thus a lower bound on the MAP decoding threshold. %
\end{conjecture}

\begin{figure*}[!t]
\normalsize 
\begin{equation} \label{eq:sc_density_evo}
\begin{split}
x^{(\ell+1)}_{i} &= %
\epsilon \sum_{g=1}^{N} A_{g,i} \left[ \sum_{h=1}^{M} A_{g,h} \left[ 1- \rho \left(1 - \sum_{p=1}^{N} A_{p,h} \left\{ \sum_{q=1}^{M} A_{p,q} \left[ 1-\rho(1-x_{q}^{(\ell)}) \right] \right\}^{k-1} \right) \right]  \right]^{k-1} %
\end{split}
\end{equation}
 \hrulefill
\vspace*{-2mm}
\end{figure*}

\section{Spatially-Coupled Counter Braids} \label{sec:SC-CBs}

We consider the ensemble $(\lambda,\rho,N,w)$ of single-layer SC-CBs \cite{ros14} (coupling the original bipartite graph), where $N$ is the coupling chain length and $w$, $1\le w\le N+1$, is a smoothing parameter \cite{kud11}. The corresponding ensemble of SC graphs is denoted by $\mathcal{G}_{\rm c}(\lambda,\rho,N,w)$. The ensemble is constructed as follows. A collection of $N$ flow-node groups are placed at positions $\{1,2,\ldots,N\}$, each containing $\kappa$ nodes of degree $k$, such that $\kappa=\frac{m_0}{N}$ (the total number of flow nodes, i.e., the number of distinct flows to be counted, is $m_0$). Furthermore, a collection of $M=N+w-1$ counter-node groups are placed at positions $\{1,2,\ldots,M\}$, each containing $\frac{\kappa R_{j} L'(1)}{R'(1)}=\frac{\kappa R_{j} k}{\gamma}$ nodes of degree $j$ (we implicitly assume that $N$ is chosen such that $\frac{\kappa R_{j} k}{\gamma}$ and also $\frac{m_0}{N}$ are integers).

The $\kappa k$ edge sockets in each group of flow and counter nodes are partitioned into $w$ equally-sized subgroups (assuming that $\frac{\kappa k}{w}$ is an integer) using a uniform random interleaver. We denote by $\mathcal{P}_{n,i}^{(f)}$ and $\mathcal{P}_{n,i}^{(c)}$ the set of flow and counter node sockets, respectively, in the $i$-th subgroup, $i=0,1,\ldots,w-1$, at position $n$, where $n=1,\ldots,N$ for flow node sockets and $n=1,\ldots,M$ for counter node sockets. The SC ensemble is constructed by adding edges that connect the sockets in $\mathcal{P}_{n,i}^{(f)}$ to the sockets in $\mathcal{P}_{n+i,i}^{(c)}$. Different graphs are obtained by different socket associations. Note that this construction leaves some sockets of the counter-node groups at the boundary unconnected and these are removed. In the following, %
we will denote  
 the coupled ensemble $\mathcal{G}_{\rm c}(\lambda,\rho,N,w)$  by the alternative notation  $\mathcal{G}_{\rm c}(k,\gamma,N,w)$. %

\subsection{Density Evolution Recursion}

Denote by $x_i^{(\ell)}$, $i=1,\ldots,M$, the error probability of an outgoing message from a flow node at position $i$ at iteration $\ell$. Note that since  there are no flow-node groups at positions $i > N$, initially $x_i^{(0)}=0$ for $N < i \leq M$.  Using the ensemble defined above, we get the recursion shown in  (\ref{eq:sc_density_evo}) at the top of the next page, where $\bm{A} = \{A_{p,q}\}$ is the $N \times M$ matrix defined by
\begin{equation} \notag
A_{p,q} = [ \bm{A}]_{p,q} = \begin{cases}
\frac{1}{w}, & \text{if $1 \leq q-p+1 \leq w$} \\
0, & \text{otherwise} \end{cases}.
\end{equation}

Note that contrary to SC-LDPC codes, for which the SC recursion contains two summations (an outer summation over $N$ terms and an inner summation over $M$ terms), the recursion for SC-CBs contains four summations (two summations over $N$ terms and two summations over $M$ terms), since the update rule for the flow nodes is different for odd and even iterations. Note that due to the band-structure of the matrix $\bm{A}$ there are only $w$ nonzero terms in all four summations. Due to this difference between odd and even iterations, SC-CBs do not fit within the general framework of coupled scalar recursions outlined in \cite{yed13}.

Let $\x^{(\ell)} = (x_{1}^{(\ell)},\ldots,x_{M}^{(\ell)})$. The BP decoding threshold of the coupled ensemble $\mathcal{G}_{\rm c}(k,\gamma,N,w)$ is defined (analogous to the uncoupled case) as ${\epsilon}_{\rm BP}^{\rm c} =  {\epsilon}_{\rm BP}^{\rm c}(k,\gamma,N,w)$, where
\begin{equation} \notag
{\epsilon}_{\rm BP}^{\rm c}(k,\gamma,N,w) 
\triangleq
\sup\left\{ \epsilon \in \mathcal{E}  \; |\;  \x^{(\infty)}(k,\gamma,\epsilon,N,w) = \bm{0} \right\}.
\end{equation}

\section{Numerical Results}
\label{sec:NumericalResults}

\begin{figure}[tbp]
\centering
\includegraphics[width=\columnwidth]{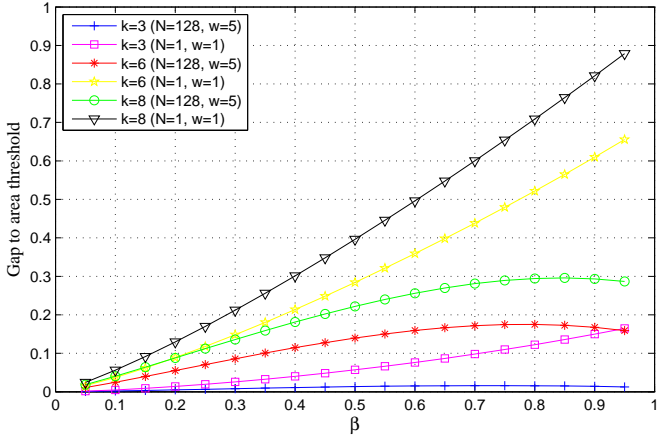}
\vspace{-4ex}
\caption{The difference $\bar{\epsilon}-{\epsilon}_{\rm BP}^{\rm c}$, where $\bar{\epsilon}$ is the area threshold from Definition~\ref{def:MAP} and ${\epsilon}_{\rm BP}^{\rm c}$ is the BP decoding threshold of SC-CBs, as a function  of $\beta$ for different values of the left-degree $k$ with $(N,w)=(128,5)$ and $(1,1)$ (uncoupled).}
\label{fig:relative_gap}
\vspace{-2ex}
\end{figure}

We have computed the BP decoding threshold ${\epsilon}_{\rm BP}^{\rm c}$ for values of $\beta$ in the range $0.05$ to $0.95$ with a step size of $0.05$ for different values of the left-degree $k$ for SC-CBs with $(N,w)=(128,5)$ and $(1,1)$ (uncoupled). %
For each value of $\beta$ (and $k$) %
we have also computed the conjectured lower bound on the MAP decoding threshold (or the area threshold) from Conjecture~\ref{th:MAP}. In Fig.~\ref{fig:relative_gap}, we plot the difference $\bar{\epsilon}-{\epsilon}_{\rm BP}^{\rm c}$, where $\bar{\epsilon}$ is the area threshold from Definition~\ref{def:MAP}, as a function of $\beta=\frac{k}{\gamma}$. As we can observe from  Fig.~\ref{fig:relative_gap}, there is a gap between the BP decoding threshold of SC-CBs (with $(N,w)=(128,5)$) and the conjectured lower bound on the MAP decoding threshold. However, the gap is significantly larger for uncoupled CBs, meaning that spatial coupling indeed improves performance. Note that the gap varies with $\beta$ %
and depends on $k$.

\section{Discussion and Future Work} \label{sec:discussion}

As shown in Section~\ref{sec:NumericalResults}, we observe threshold improvement but not threshold saturation to the area threshold by coupling the original bipartite graph (see Section~\ref{sec:SC-CBs}). We remark that coupling the equivalent graph representation mentioned in Section~\ref{sec:AsymptoticAnalysis} %
gives exactly the same coupled DE recursion in (\ref{eq:sc_density_evo}). However, interestingly, if we consider the \textit{coupled version} of (\ref{eq:DECounter2})-(\ref{eq:g}), i.e., we substitute $y$ and $x$ in (\ref{eq:f}) and (\ref{eq:g}), respectively, by an average over spatial positions, we do indeed observe threshold saturation to the area threshold. The DE recursion derived below corresponds to the coupled version of (\ref{eq:DECounter2})-(\ref{eq:g}), split into two steps.

Let $x_i^{(\ell)}$ and $y_i^{(\ell)}$, $i=1,\ldots,M$, denote the output message error probability at a flow and counter node, respectively, at the $\ell$-th iteration at coupling chain position $i$. As before, we initialize $x_i^{(0)}=0$ for $N < i \leq M$. %
Now, define the following DE  equations at iteration $\ell$,
\begin{align}
y_i^{(\ell)} &= \begin{cases}
1- \rho\left( 1- \frac{1}{w} \sum_{j=0}^{\min(i-1,w-1)} x_{i-j}^{(\ell-1)} \right), & \text{if $\ell$ is odd} \\[1.0ex]
1- \rho\left( 1- x_{i}^{(\ell-1)} \right), & \text{if $\ell$ is even} \end{cases}
\notag \\
x_i^{(\ell)} &= \begin{cases}
\left( y_i^{(\ell)} \right)^{k-1},  & \text{if $\ell$ is odd} \\[1.0ex]
\epsilon \left( \frac{1}{w} \sum_{j=0}^{\min(M-i,w-1)} y_{i+j}^{(\ell)} \right)^{k-1},  & \text{if $\ell$ is even} \end{cases}. \notag
\end{align}
With this modified DE, threshold saturation to the area threshold which (from Conjecture~\ref{th:MAP}) is a lower bound on the MAP decoding threshold, can be proved using the potential function framework by Yedla \emph{et al.} outlined in \cite{yed13}. Note that this DE is characterized by an average only for odd and even iterations for the counter node and flow node updates, respectively. However, when coupling the original (or the equivalent) graph in the standard way, the average appears for all iterations (hence the four summations in (\ref{eq:sc_density_evo})). This effect, which is due to the fact that, as opposed to LDPC codes, the flow node update is different for odd and even iterations, seems to be the responsible for the lack of threshold saturation. 

A question that remains open is whether the DE equations above correspond to a physical system, i.e., whether threshold saturation can be achieved with an alternative coupling. Future research also includes a proof of Conjecture~1, the analysis of CBs with more layers, and exploring the connection with compressed sensing.

\section*{Acknowledgment}
The authors are grateful to Henry D. Pfister for useful discussions and insightful comments.

\balance


\end{document}